\documentclass[preprint,12pt]{elsarticle}

\usepackage{amssymb}
\usepackage{amsmath}
\usepackage{hyperref}
\usepackage{listings}
\usepackage{xcolor}

\definecolor{keywords}{RGB}{0,125,0}
\definecolor{comments}{RGB}{0,100,113}
\definecolor{red}{RGB}{160,0,0}
\definecolor{green}{RGB}{0,150,0}

\journal{Nuclear Physics B}

\begin{document}

\begin{frontmatter}



\title{Shaping circuit for improving linearity, bandwidth, and dynamic range in ToT-based detectors}


\author[BUW]{J. Peña-Rodríguez}
\author[BUW]{J. Förtsch}
\author[BUW]{C. Pauly}
\author[BUW]{K.-H. Kampert}

\affiliation[BUW]{organization={Bergische Universität Wuppertal, Fakultät für Mathematik und Naturwissenschaften, Gaußstraße 20, 42119 Wuppertal, Germany}
}

\begin{abstract}
The quantitative measurement of energy deposits in particle detectors, particularly in calorimeters,
is usually accomplished with the help of Analog-to-Digital converters (ADCs) due to their precision, wide measurement range, and good linearity. However, drawbacks such as power consumption, data volume, and bandwidth limit their use in the next generation of high-energy physics experiments. Time-over-threshold (ToT) systems offer simplicity, low power consumption, easy integrability, and wide bandwidth, but they lack precision, linearity, and dynamic range. In this work, we propose a shaper circuit that improves the weaknesses of ToT systems without sacrificing performance. We simulated and implemented the concept in the readout system of the Ring Imaging Cherenkov detector of the Compressed Baryonic Matter experiment at FAIR.

\end{abstract}

\begin{graphicalabstract}
\includegraphics[width=15cm]{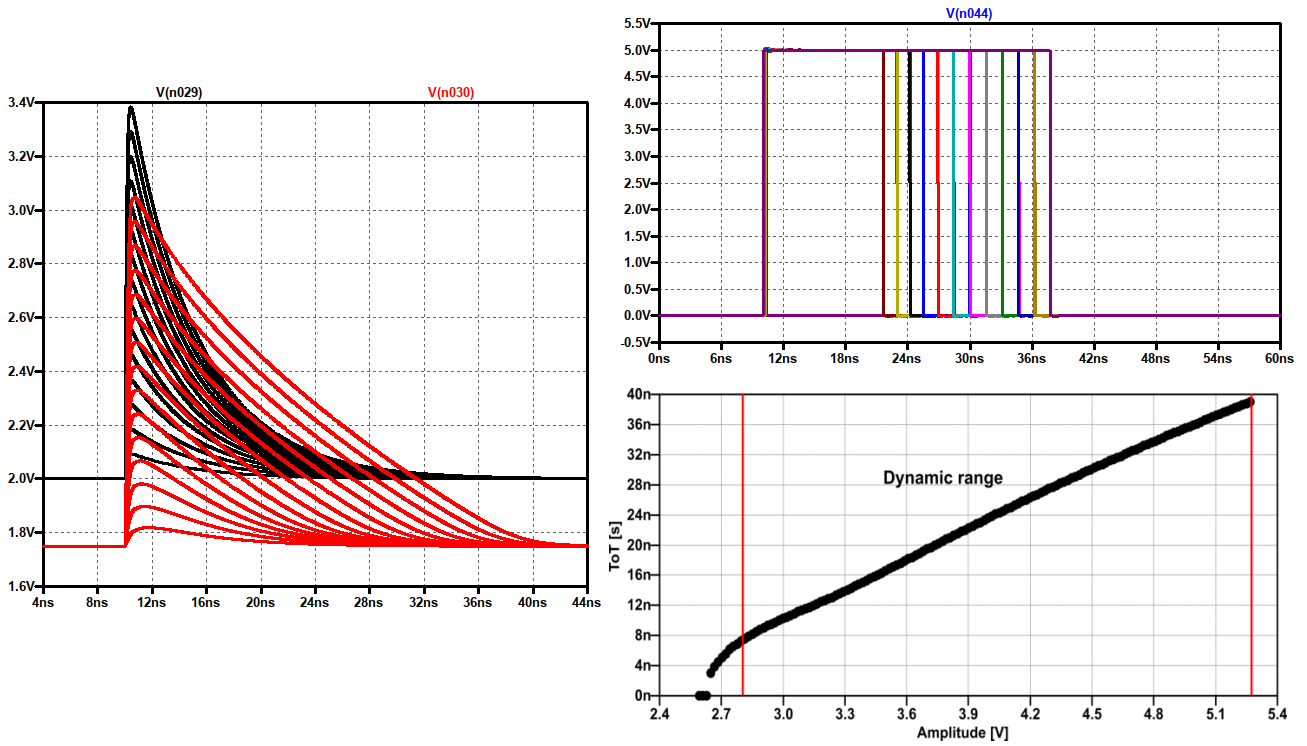}
\end{graphicalabstract}

\begin{highlights}
\item A conditioning circuit to improve the bandwidth, linearity, and dynamic range of particle detector readouts based on Time-over-Threshold.
\item Validation of the proposed concept by implementing it in the readout electronics of the Ring Imaging Cherenkov detector of the Compressed Baryonic Matter experiment.
\item The simplicity of the proposed concept reduces power consumption and improves the integrability of the circuit per channel, enhancing its implementation in Application Specific Integrated Circuits (ASICs) for particle detectors.

\end{highlights}


\end{frontmatter}


\section{Introduction}
\label{intro}

Most particle detectors use photosensors, such as silicon photomultipliers (SiPMs) or photomultiplier tubes (PMTs) to record the photons produced by the interaction of particles with the detector bulk material ionization, Cherenkov radiation, or scintillation processes. SiPMs comprise thousands of avalanche photodiodes (APDs) that are arranged parallel and biased above the breakdown voltage. When a photon interacts with one of these APDs, an electron-hole pair is created, which triggers an avalanche of charge carriers in the gain region due to the electric field effect. The SiPM anode conducts the avalanche current toward an external electronics for readout. The principle of functioning of PMTs is different from that of SiPMs. A PMT contains a photocathode, multiple dynodes, and an anode inside a vacuum tube. When a photon impinges on the photocathode surface, the photoelectric effect knocks out a primary electron. The primary electron then accelerates toward the first dynode,  knocking out secondary electrons, which are accelerated again toward the next dynode. This process occurs along all the dynodes up to the anode, multiplying the number of secondary electrons, and defining the PMT gain. The output charge signal has a similar waveform in both SiPMs and PMTs. The signal rises to a maximum amplitude and then decays exponentially until it reaches steady state.

Information about the physical interaction within the detector, particularly the deposited energy, can be obtained by measuring the light yield. This is because, in both SiPMs and PMTs, the signal amplification is - in good approximation - linear over a large dynamic range so that the electrical output signal (or charge) scales with the number of detected photons \cite{knoll2010radiation}.

Most particle detector readouts use sampling Analog-to-Digital converters to quantize the
output signal charge and obtain a measure of the deposited energy. This approach requires buffer storage, which depends on the sampling frequency. The higher the sampling frequency, the more precise the energy estimation, and the larger the required buffer size. These requirements create a bottleneck for the expected latency of future high-energy physics  experiments cite{Aad2022,Brning2022,Petyt2018}.

Time-over-threshold (ToT) readouts measure the time difference between the leading and trailing edges of a signal, typically the logic signal output of a discriminator. This methodology significantly reduces the required buffer size, circuit complexity, and power consumption, thereby improving circuit integration and counting rates. The main drawbacks of ToT systems are the degradation of energy resolution, dynamic range, and linearity \cite{Gaudin2020}. ToT systems for measuring particle energy have been implemented in Application-Specific Integrated Circuits (ASICs) \cite{Orita_2018, Gomez2021, Sanchez2022}. 

\begin{figure}[h!]
    \centering
    \includegraphics[width=11cm]{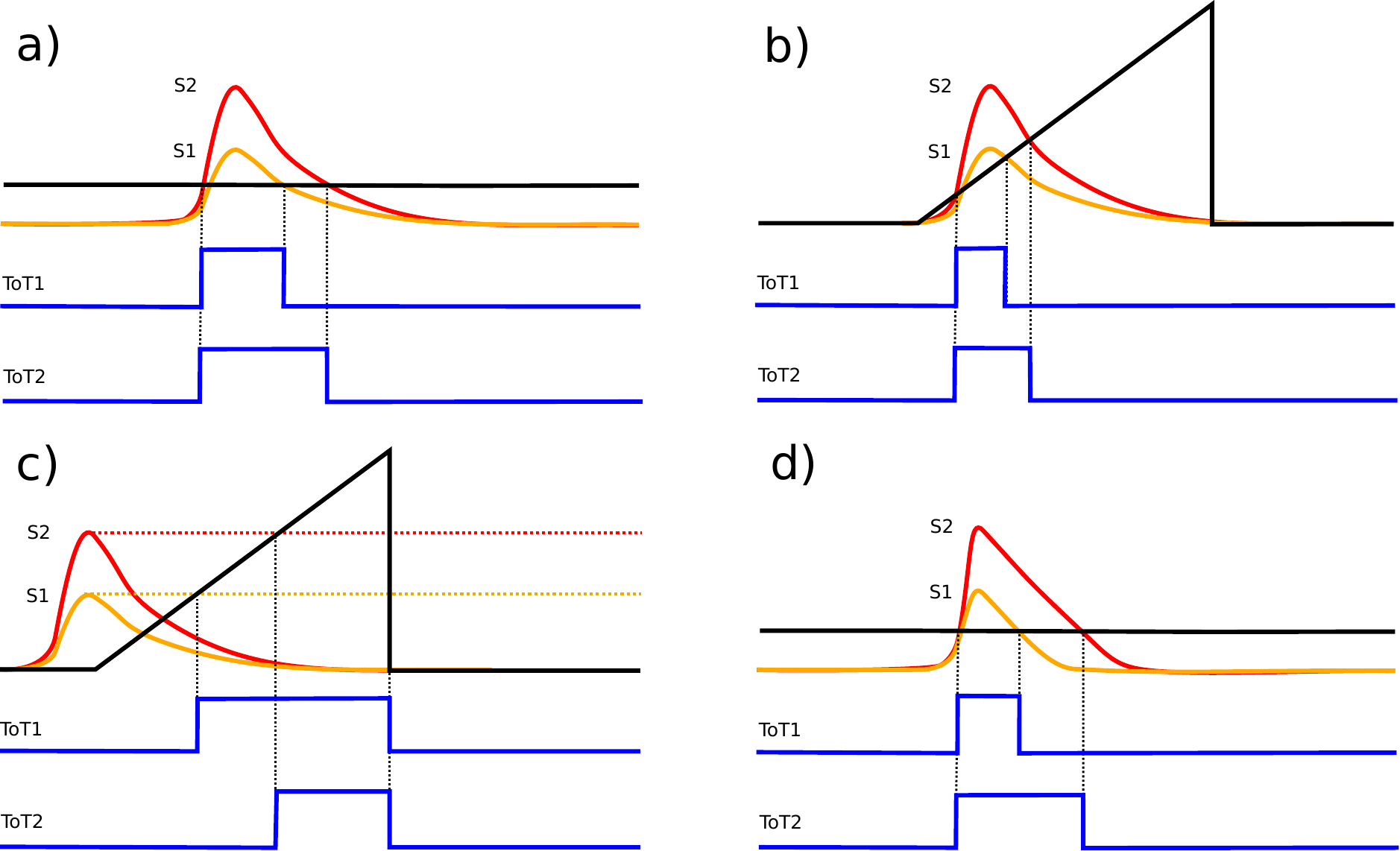}
    \caption{Different approaches to time-over-threshold measurement: (a) ToT based on constant threshold comparison, (b) ToT based on dynamic threshold, (c) ToT based on peak holding/dynamic threshold, and (d) ToT based on pulse trailing edge linearization.}
    \label{fig:tot_principles}
\end{figure}

There are different approaches to implementing the analog signal discrimination required for ToT measurement. One approach is to compare the detector pulse signal with a constant threshold and another one is to compare it a dynamic threshold \cite{Shimazoe2012}. A third approach is to hold the maximum pulse amplitude and compare it with a dynamic threshold \cite{Gomez2021, Ahmad2018}, as shown in Fig.\ \ref{fig:tot_principles}. However, these approaches have drawbacks, such as implementation complexity, power consumption per channel, nonlinearities, and limited detection rate capacity. Table \ref{tab:tot_systems} summarizes the characteristics of ToT methods.

\begin{table}[h!]
\footnotesize
    \centering
    \begin{tabular}{lcccc}  \hline
     \textbf{ToT method}    & \textbf{Linearity} & \textbf{Range} & \textbf{Power consumption} & \textbf{Counting rate} \\ \hline
     constant threshold   & low & low & low & high \\ 
     dynamic threshold    & medium & medium & medium  & medium  \\ 
     D$\&$H + dynamic th.   & high & high& high  & low \\ 
     Trailing edge shaping   & high & medium & low & high \\ \hline 
    \end{tabular}
    \caption{Characteristics of different ToT methods. }
    \label{tab:tot_systems}
\end{table}

In this work, we present a simple signal shaping method that improves the ToT determination and features good linearity, a large bandwidth, low power consumption, and high integrability. We characterized the ToT  method was under controlled conditions and tested it in the DiRICH readout system of the Ring Cherenkov detector of the Compressed Baryonic Matter experiment at FAIR \cite{AdamczewskiMusch2017}.

\section{Pulse trailing edge linearization}
\label{sec::circuit}

The fundamental problem with single-threshold ToT converters in photosensor readouts is the shape of the photosensor signal. As the amplitude (or charge) of the signal increases well above the detection threshold, the trailing edge of the output signal decays exponentially and the ToT becomes saturated, causing a degradation in energy resolution \cite{Gaudin2020}. Linearizing the  trailing edge of the signal offers an optimal solution for obtaining a linear energy/ToT ratio at a high detection bandwidth using a single-threshold ToT system.

\begin{figure}[h!]
    \centering
    \includegraphics[width=5.5cm]{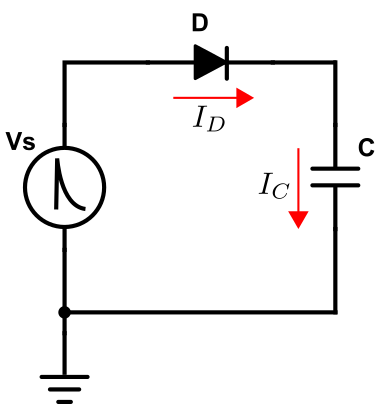}
    \caption{Capacitor charging cycle. The red arrow represents the current flowing from the photosensor to the capacitor when the diode is forward polarized.}
    \label{fig:charge}
\end{figure}

In the ToT approach proposed in this work, the photosensor signal ($V_s$) is modified by a shaping circuit consisting of a high-bandwidth diode ($D$) connected in series to a capacitor (C) as illustrated in Fig.\ \ref{fig:charge}. 

When the photosensor generates a signal, a current flows from the photosensor through the diode and charges the capacitor until it reaches $V_{max}$. Once the maximum value is reached, the photosensor signal decreases rapidly, which reverse-biases the diode and slowly discharges the capacitor with an approximately constant current \cite{Roy2018}. 


\subsection{Capacitor charging}

During the charging cycle, the diode voltage is equal to the difference between the photosensor voltage $V_s$ and the capacitor voltage $V_C$,

\begin{equation}
    V_C = V_s - V_D \: .
\label{eq:vd}
\end{equation}

In forward polarization, $V_D$ is constant. This means that $V_C$ follows the same wave shape as $V_s$, but with the value of $V_D$ subtracted.

\subsection{Capacitor discharging}

\begin{figure}[h!]
    \centering
    \includegraphics[width=7cm]{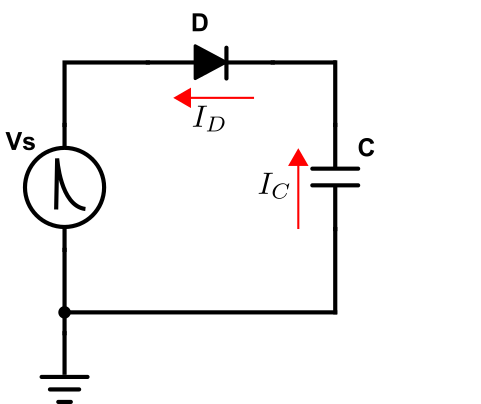}
    \caption{Capacitor discharging cycle. The red arrows show the current flowing from the capacitor to the photosensor when the diode is reverse polarized.}
    \label{fig:discharge}
\end{figure}

Once $V_s$ reaches the maximum $V_{max}$, it rapidly decreases to zero and the diode $D$ is reverse biased because $V_C > V_s$ as shown in Fig.\ \ref{fig:discharge}. The diode voltage then becomes $V_D = V_C$.

The circuit differential equation is

\begin{equation}
    I_s \left( e^{V_C/\eta V_T} - 1 \right) = C \frac{dV}{dt},
\end{equation}

where $I_s$ is the reverse-bias saturation current, $\eta$ is the ideality factor, which typically varies from 1 to 2), and $V_T = kT/q$ is the thermal voltage related to the temperature of the p-n junction of the diode. The model solution is:

\begin{equation}
    V_C(t)=V_T \ln \left[ 1 - \left(1 - e^{V_s/V_T} \right)e^{- \alpha t} \right],
\end{equation}
where

\begin{equation}
    \alpha = \frac{I_s}{CV_T}.
\end{equation}
Due to $V_s \gg V_T$, then $e^{V_s/V_T} \gg 1$, then

\begin{equation}
    V_C(t)=V_T \ln \left[ 1 + e^{V_s/V_T - \alpha t} \right].
\end{equation}

For values where $e^{V_s/V_T - \alpha t} \gg 1$, the expression reduces to:

\begin{equation}
    V_C(t)=V_T \ln \left[ e^{V_s/V_T - \alpha t} \right] = V_{max} - \frac{Is}{C} t.
\end{equation}
The capacitor discharges from $V_{max}$ with a linear ratio of $I_s/C$. The discharge time is

\begin{equation}
    t = \frac{C [V_{max} - V_C(t)]}{I_s}.
\end{equation}

\section{Spice simulations}

We performed Spice simulations to evaluate the circuit performance. Two fast diodes were tested, the 1N5817 and the SMS7621, as well as the BFU760 transistor operating as a diode with base-emitter terminals short-circuited. High-frequency transistors offer faster switching characteristics than regular diodes in low-power and high-speed applications, particularly when used in integrated circuits. The diode parameters are listed in Table \ref{tab:parameter}.

\begin{table}[h!]
\centering
\begin{tabular}{clccc} \hline
 \textbf{Parameter} &\textbf{Description} & \textbf{Units} & \textbf{SMS7621}  & \textbf{1N5817}  \\\hline
 $I_S$ & Saturation current   & A & $4\times 10^{-8}$ &  $31.7\times 10^{-6}$\\ 
 $R_S$ & Parasitic resistance & $\Omega$ & 12 & 0.051  \\
 $N$     & Emission coefficient& - & 1.05 & 1.373  \\
 $TT$    & Transit time & s  & $1\times 10^{-11}$ & $1\times 10^{-10}$  \\ 
 $C_{JO}$   & Zero-bias junction capacitance & pF & 0.1 & 190  \\
 $M$     & Junction grading coefficient & - & 0.35 & 0.3  \\
 $E_G$    & Activation energy & eV & 0.69 & 0.69  \\ 
 $X_{TI}$   & $I_S$ temperature exponent & - & 2 & 2  \\
 $F_C$    & Forward bias depletion & - & 0.5 & 0.5  \\
        & capacitance coefficient &  &  &  \\
 $B_V$    & Reverse breakdown voltage & V & 3 & 22   \\ 
 $I_{BV}$   & Reverse breakdown current & A & $1\times 10^{-5}$ & 1  \\
 $V_J$    & Junction potential & V & 0.51 & 0.35  \\\hline
\end{tabular}
\caption{Spice parameters of the SMS7621 and 1N5817 diodes. }
\label{tab:parameter}
\end{table}

\begin{figure}[h!]
    \centering
    \includegraphics[width=7cm]{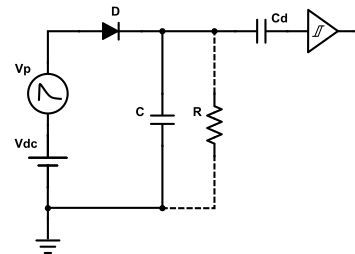}
    \caption{Simulated circuit. The baseline of the photosensor signal $V_p$ is $V_{dc}$. The capacitor $C_d$ decouples the DC component of the shaped signal before it reaches the discriminator. $R$ represents the resistive load.}
    \label{fig:spice_circuit}
\end{figure}

The simulated circuit is shown in Fig. \ref{fig:spice_circuit}. The amplitude of the photosensor signal $V_p$ ranges from 0.1\,V to 3\,V in 0.1\,V increments, with a rise time of 100\,ps and a decay time of 5\,ns. A constant bias signal offset of 2.5\,V is added to forward bias the diode and accommodate small signal amplitudes. The capacitor $C_d$ decouples the DC offset prior to signal discrimination. The comparator creates a square signal by comparing the circuit output to a 100\,mV threshold. We took measurements of the input amplitude and the output Time-over-Threshold.

\subsection{Circuit performance without resistive load}

\begin{figure}[h!]
    \centering
    \includegraphics[width=6.7cm]{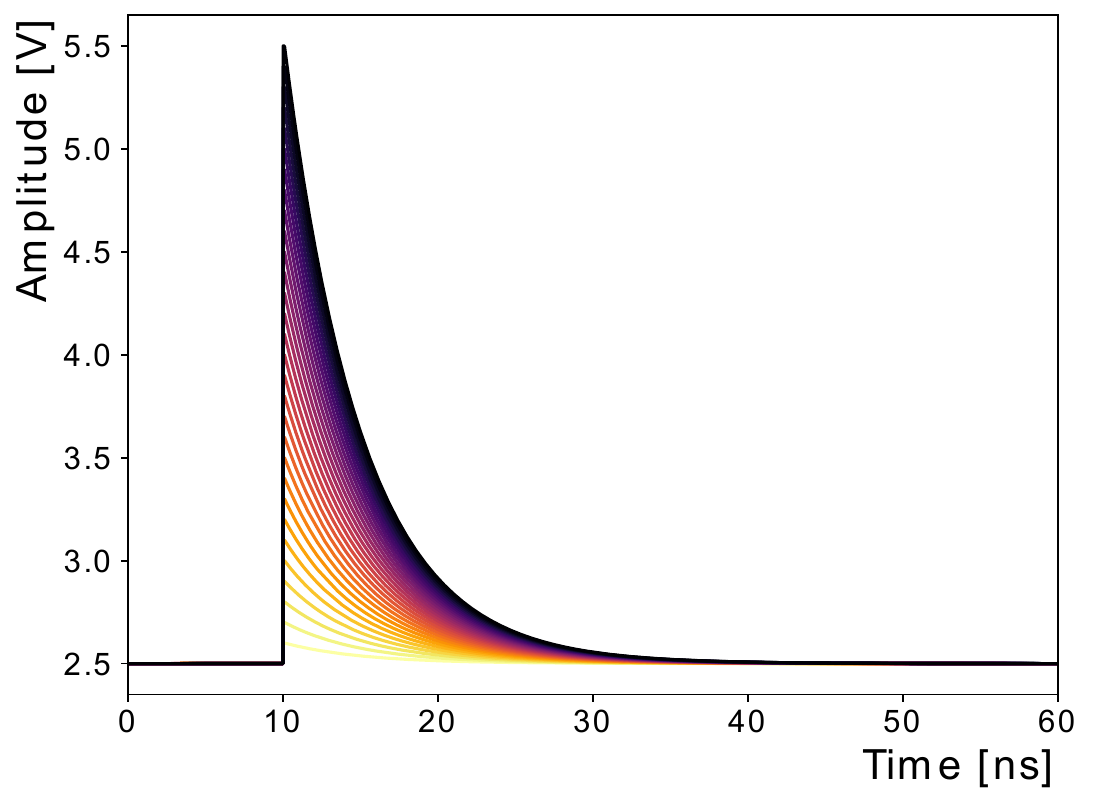}
    \includegraphics[width=6.7cm]{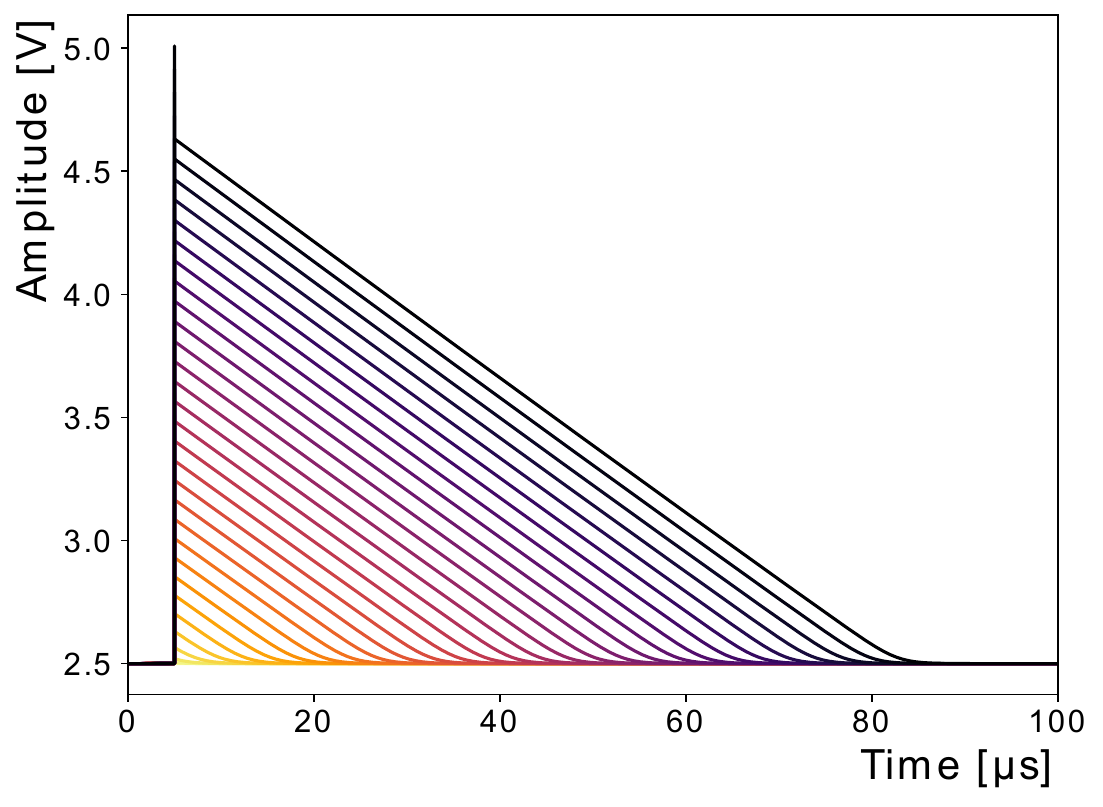}
    \caption{The left graph shows the simulated photosensor signal with a peak amplitude ranging from 0.1\,V to 3\,V in 0.1\,V increments. The output signal of the trailing edge shaper has a 1\,nF shaping capacitance and no resistive load (right). The discharging rate is constant rate of 31.7\,mV/\textmu s.}
    \label{fig:no_load}
\end{figure}

The first step was to evaluate the performance of the circuit under no-load conditions. We set the shaper capacitance to 1\,nF. Figure \ref{fig:no_load} shows the photosensor signals for different peak amplitudes (left) and the corresponding shaped signals using the 1N5817 diode (right). The photosensor signal exhibits typical exponential decay ($\tau = 5$\,ns), while the shaped signal decreases at a constant rate of 31.7\,mV/\textmu s. 

\begin{figure}[h!]
    \centering
    \includegraphics[width=7cm]{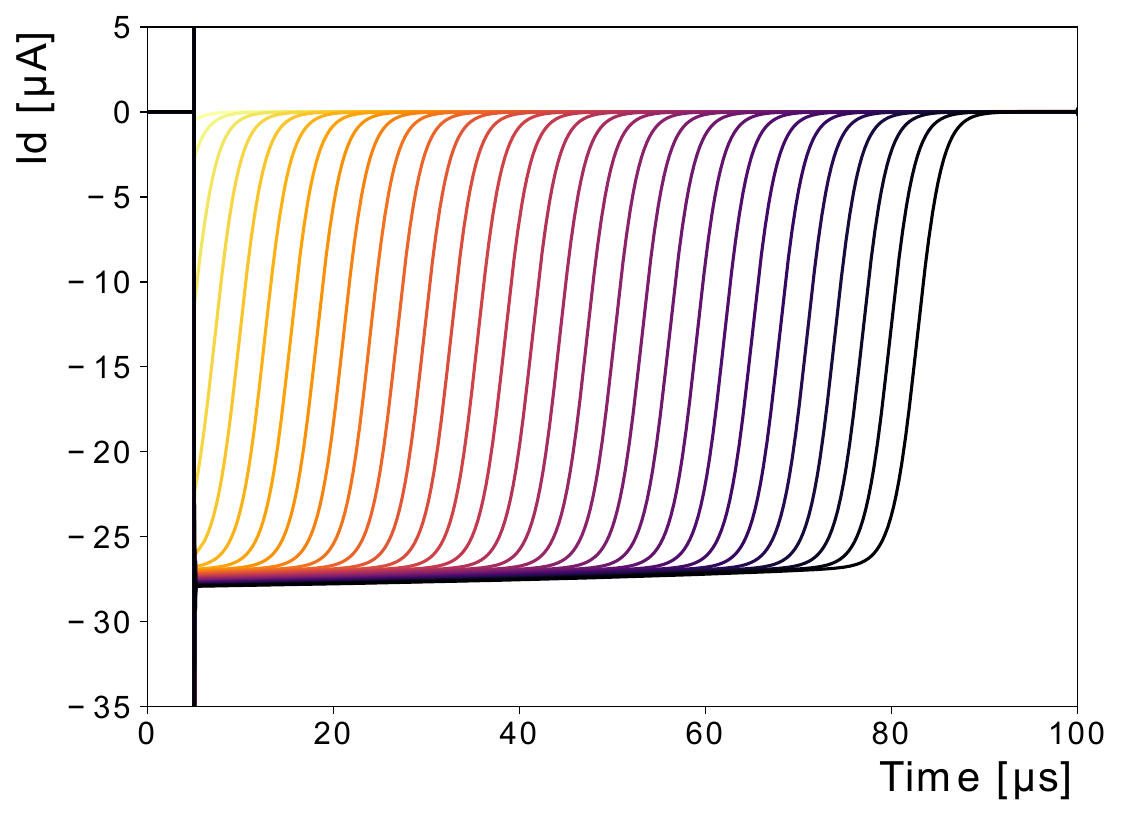}
    \caption{Diode current during the discharge cycle for different photosensor signals. The discharging current remains constant at around the diode saturation current of 28\,\textmu A.}
    \label{fig:no_load_Ic}
\end{figure}

As shown in Fig.\ \ref{fig:no_load_Ic}, the capacitor discharges at a constant current set by the diode. The discharge current remains close to the diode saturation current (31.7\,\textmu A for the 1N5817 diode), and quickly drops to zero when the capacitor is almost discharged, reaching the baseline voltage of 2.5\,V. 

\subsection{Circuit performance with resistive load}

\begin{figure}[h!]
    \centering
    \includegraphics[width=7.2cm]{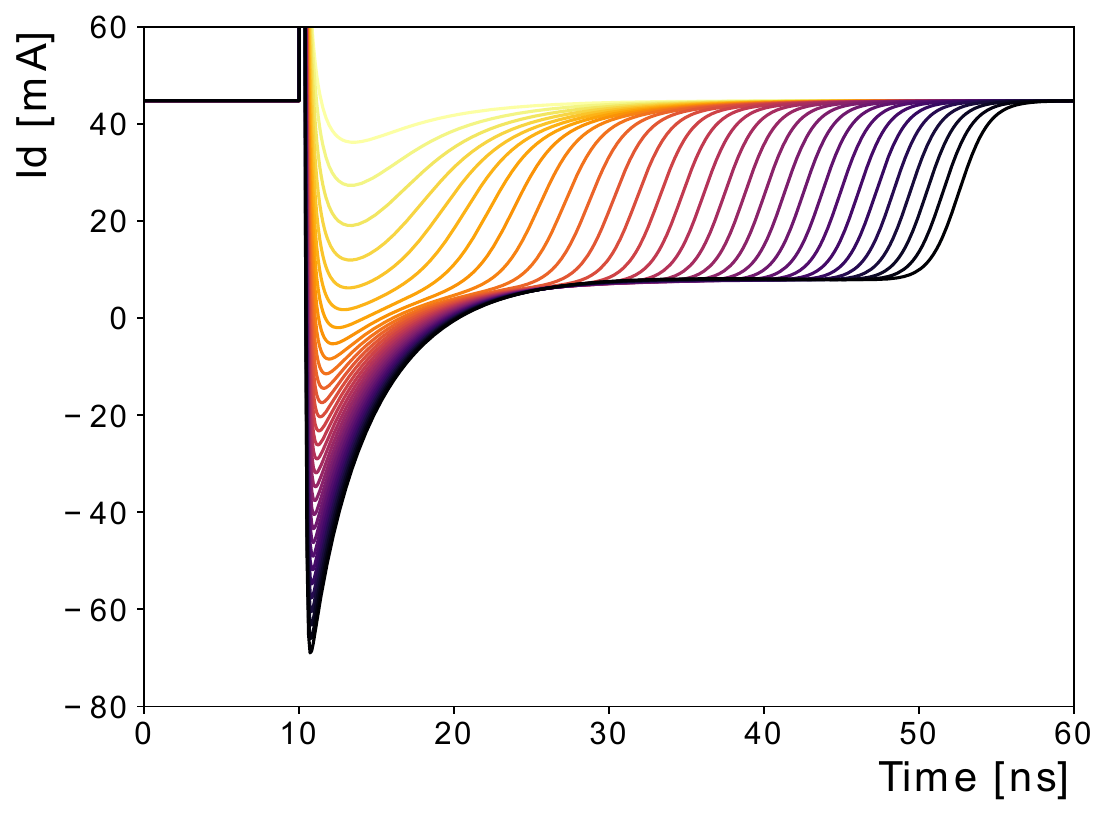}
    \caption{Diode current during the discharging cycle including a resistive load of 50\,$\Omega$.}
    \label{fig:curr_load}
\end{figure}

In this step, we evaluate the performance of the circuit after adding a 50\,$\Omega$ resistive load to reduce the discharge time of the capacitor. Figure \ref{fig:curr_load} shows how the diode current is shifts positively from 0\,mA to a new reference point of approximately 45\,mA. The resistive load continuously drains current from $V_{dc}$, setting a current offset to the system. The diode current behaves exponentially during the photosensor trailing edge ($<25$\,ns) before reaching the steady state where the diode current becomes constant and the discharge of the capacitor linear.  The inverse diode current increases to about 38\,mA with respect to the new reference point. The lower limit of the product $RC$ is the decay constant $\tau$ of the photosensor signal. Based on the simulation results, we recommend $RC >= 10 \tau$.

\begin{figure}[h!]
    \centering
    \includegraphics[width=6.7cm]{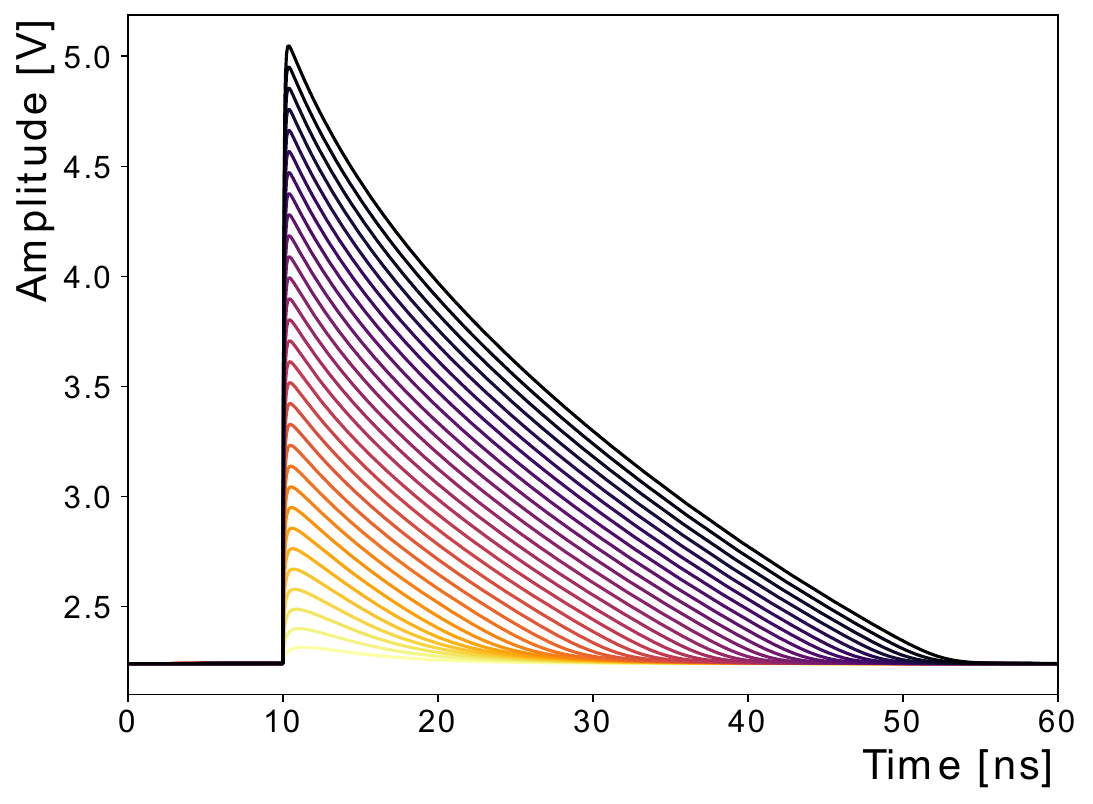}
    \includegraphics[width=6.7cm]{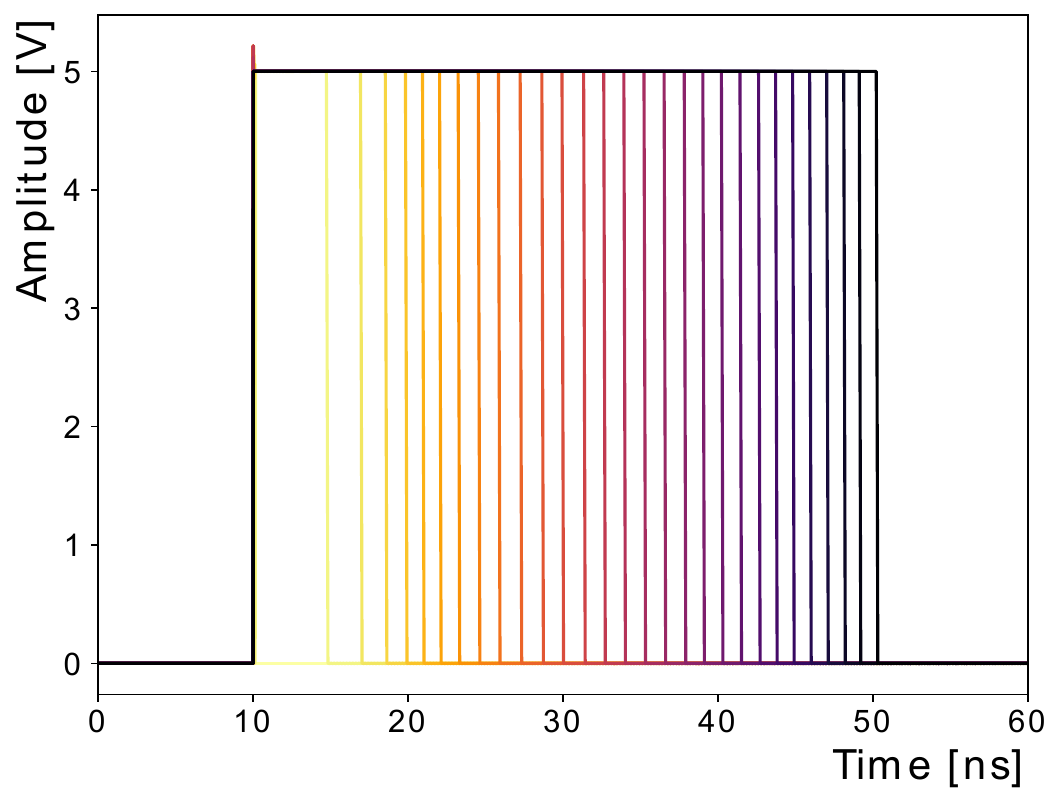}
    \caption{Output of the signal shaper with a resistive load of 50\,$\Omega$ (left). Output signal of the discrimination system (right).}
    \label{fig:load}
\end{figure}

Figure \ref{fig:load} shows the shaper output (left) and the discriminator output (right) for different photosensor signals with amplitudes ranging from 0.1\,V to 3\,V in 0.1\,V increments.  Using the resistive load, the capacitor discharges in tens of nanoseconds, which is three orders of magnitude faster than without the resistive load (tens of microseconds). The ToT value of the discriminator output increases proportionally with the shaper output amplitude. 

\begin{figure}[h!]
    \centering
    \includegraphics[width=6.5cm]{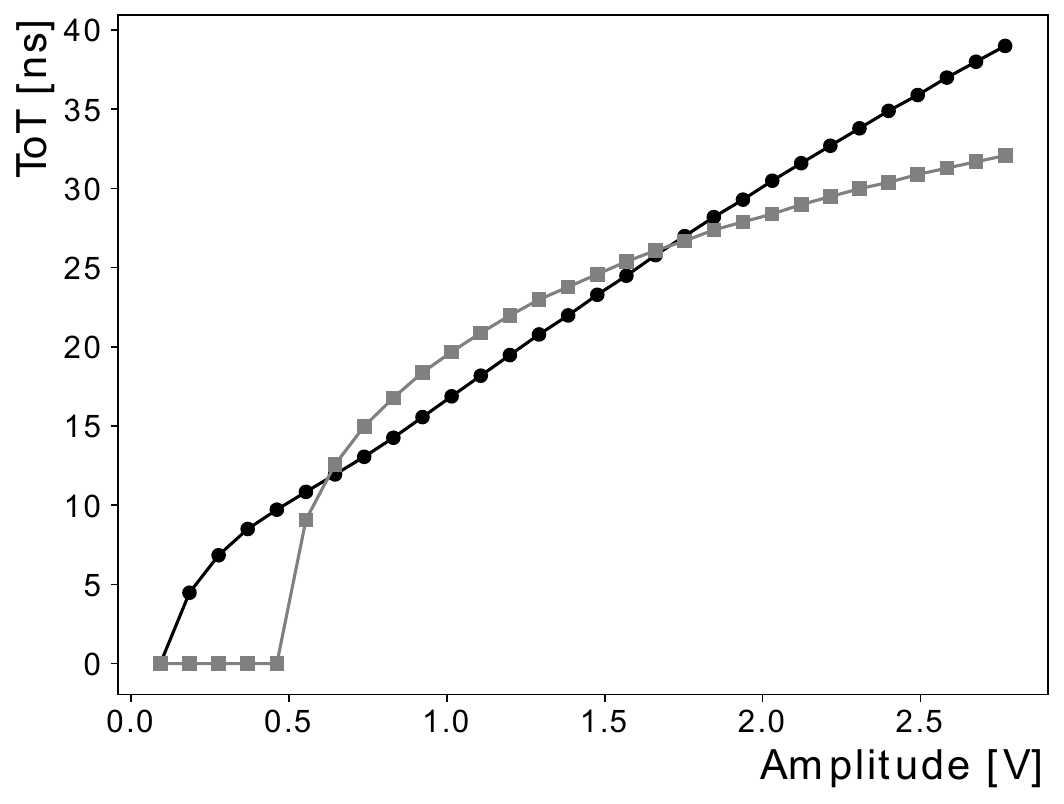}
    \includegraphics[width=6.5cm]{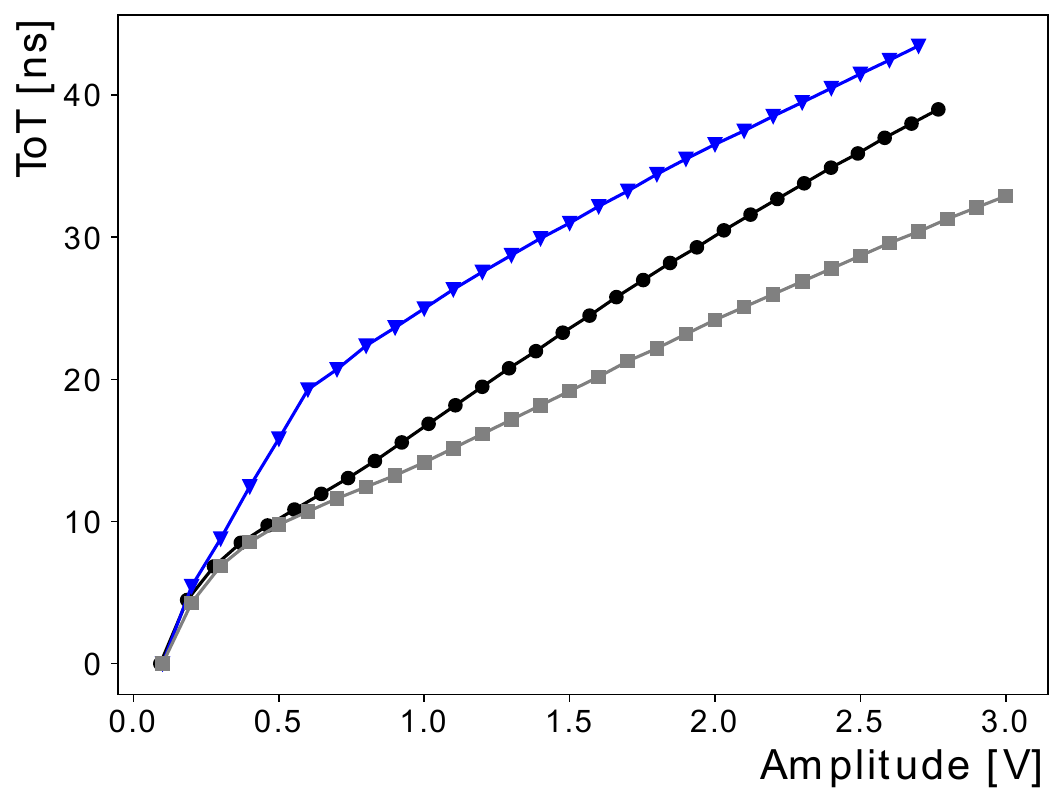}
    \caption{ToT linearity of the shaper circuit using the 1N5817 and SMS7621 diodes with a shaping capacitance of 1\,nF and a resistive load of 50\,$\Omega$ (left). Shaper circuit performance for two diodes (1N5817 and SMS7621) and the BFU760 transistor after frequency bandwidth broadening.}
    \label{fig:compare}
\end{figure}

We tested the linearity of the circuit with two diodes (1N5817 and SMS7621) and a BFU760 transistor. The shaper capacitance was 1\,nF and the resistive load was 50\,$\Omega$. The characteristic amplitude vs.\ ToT of the 1N5817 diode shows a linear tendency, whereas for the SMS7621 diode, the curve tends to saturate as the amplitude increases, as shown in Fig.\ \ref{fig:compare} (left). This effect is caused by a combination of the diode parameters (junction capacitance $C_{JO}$ and parasitic resistance $R_S$) and the shaper components ($C$ and $R$). The small junction capacitance and the relatively high parasitic resistance of the SMS7621 ($C_{JO}=$ 0.1\,pF and $R_S=12$\,$\Omega$) suppress frequency components of the shaped signal, whereas the 1N5817 diode preserves them ($C_{JO}=$ 190\,pF and $R_S=0.051$\,$\Omega$). On the other hand, the BFU760 transistor has a base-collector zero-bias capacitance ($C_{JC}$) of 21.67\,fF and a collector resistance ($R_{C}$) of 50.31\,$\Omega$. Again, a low junction capacitance combined with a high parasitic resistance degrades the shaper performance.

\begin{figure}[h!]
    \centering
    \includegraphics[width=8.5cm]{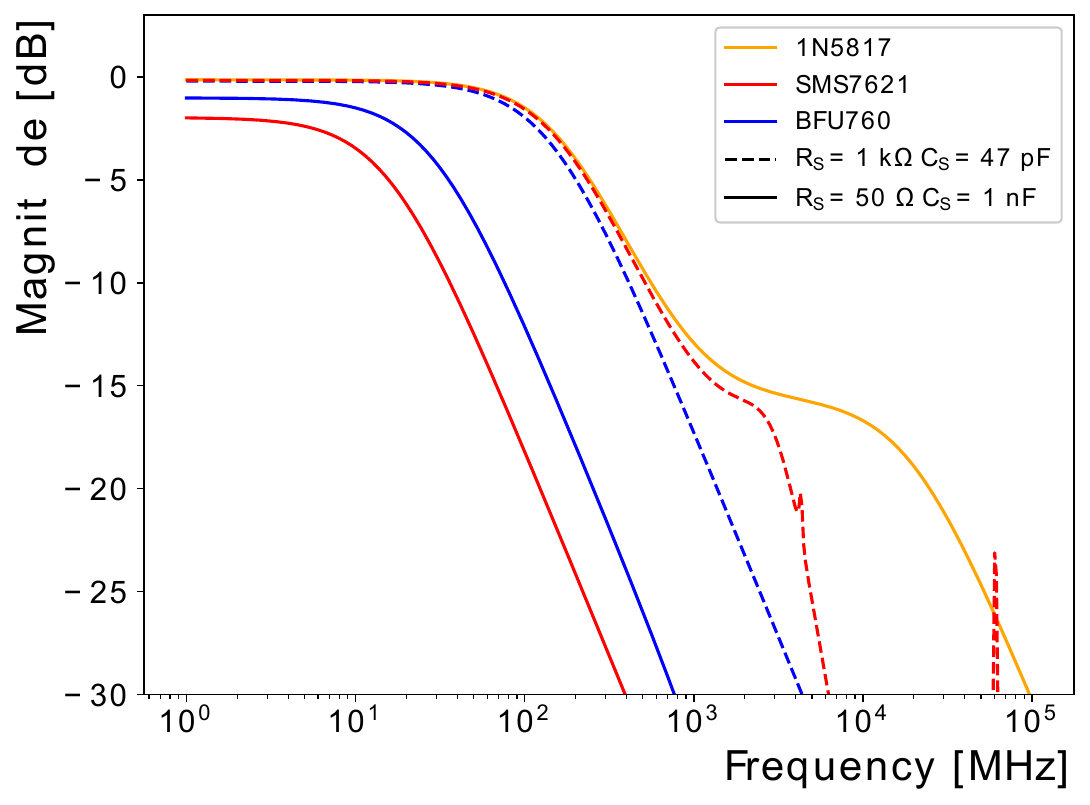}
    \caption{Frequency response of the shaper circuit for the 1N5817 and SMS7621 diodes, and the BFU760 transistor using two capacitance and resistive load configurations. The solid line represents the circuit performance with $R=50$\,$\Omega$ and $C=1$\,nF, while the dashed line with $R=1$\,k$\Omega$ and $C=47$\,pF.}
    \label{fig:bode}
\end{figure}

To improve the frequency response without sacrificing discharge time, we decreased the shaper capacitance to 47 pF and increased the resistive load to 1 k$\Omega$. Figure \ref{fig:bode} shows the frequency response of the shaper circuit with the 1N5817 diode, the SMS7621 diode and the BFU760 transistor. Figure\,\ref{fig:compare} (right) shows how the circuit linearity improves for the SMS7621 diode and the BFU760 transistor after the bandwidth broadening. This analysis revelaed that the junction capacitance and the parasitic resistance of the shaper diode are crucial for circuit performance. 

\begin{figure}[h!]
    \centering
    \includegraphics[width=6.5cm]{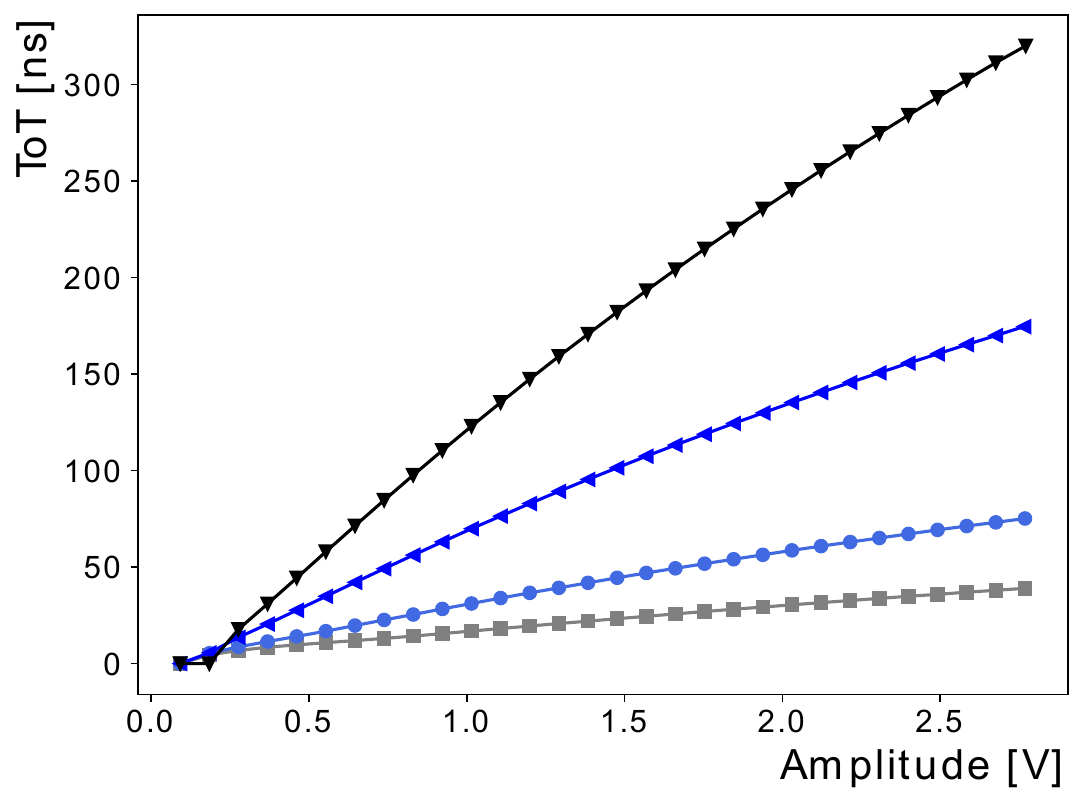}
    \includegraphics[width=6.5cm]{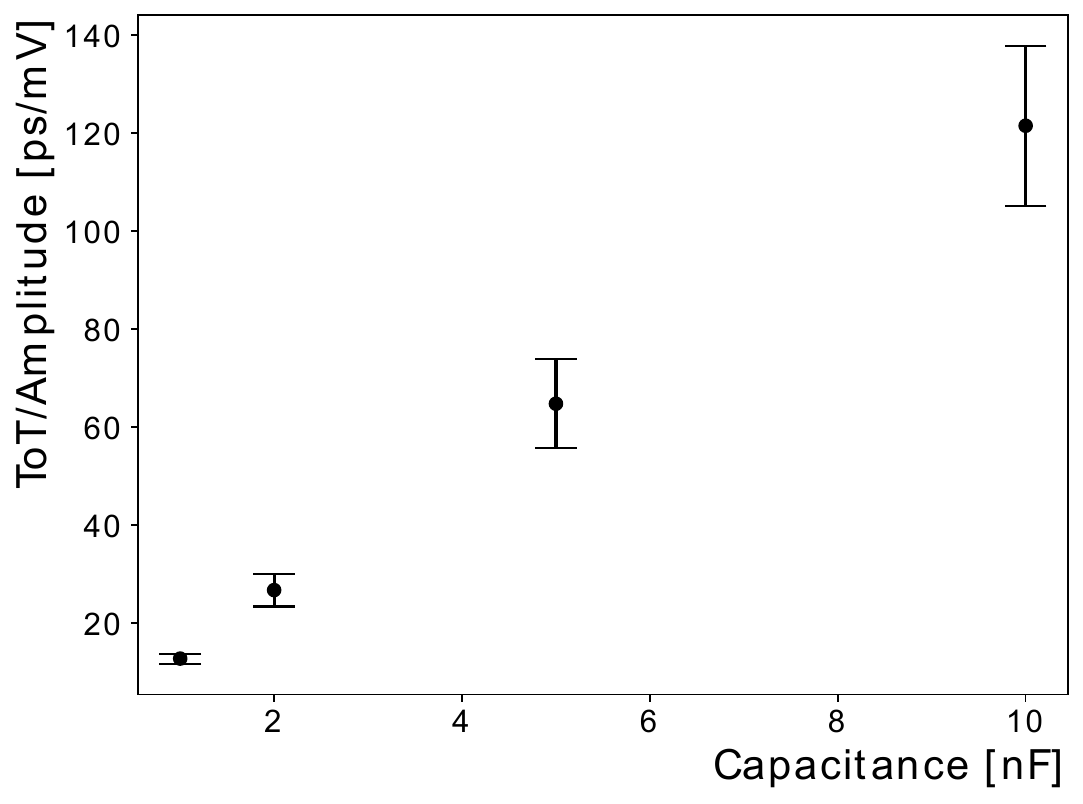}
    \caption{Simulations of the ToT dependency on the shaper capacitance. Amplitude vs.\ ToT characteristic for 1\,nF, 2\,nF, 5\,nF, and 10\,nF (left). $\Delta_{ToT}/\Delta_{Amp}$ variation depending on the shaper capacitance (right).}
    \label{fig:capa}
\end{figure}

We tested the performance of the circuit (with a 1N5817 diode) by varying the shaper capacitor. As shown in Fig.\,\ref{fig:capa}, we found that the slope $\Delta_{ToT}/\Delta_{Amp}$ increases with the shaper capacitance. However, the ToT resolution improves by decreasing the capacitance. An improvement in ToT resolution drives an improvement in charge/amplitude resolution as well as energy resolution, e.g.\ in a calorimeter.

\begin{equation}
    \frac{\sigma_{ToT}}{\Delta_{ToT}} \sim \frac{\sigma_Q}{\Delta_Q} \sim \frac{\sigma_E}{\Delta_E},
\end{equation}

\section{Circuit temperature dependency}

The potential temperature dependency must be taken into account in the circuit implementation. The diode saturation current $I_S$ is the sum of the generation current $I_{gen}$, which is caused by the thermal generation of charge carriers (electrons/holes) within the diode depletion zone, and the diffusion current $I_{diff}$, which is caused by the minority of carriers in the regions n or p diffusing across the depletion zone due to the effect of the junction field. Although $I_S$ is independent of voltage, it depends on temperature due to its thermally stimulated contributors.

The generation current is defined as

\begin{equation}
    I_{gen}(T) = e^{-qV_J/2kT},
\end{equation}
where, $V_J$ is the junction potential, $q$ the electron charge, $k$ is the Boltzmann constant, and $T$ is the absolute temperature.
The diffusion current is

\begin{equation}
    I_{diff}(T) = e^{-qV_J/kT},
\end{equation}

\begin{figure}[h!]
    \centering
    \includegraphics[width=8cm]{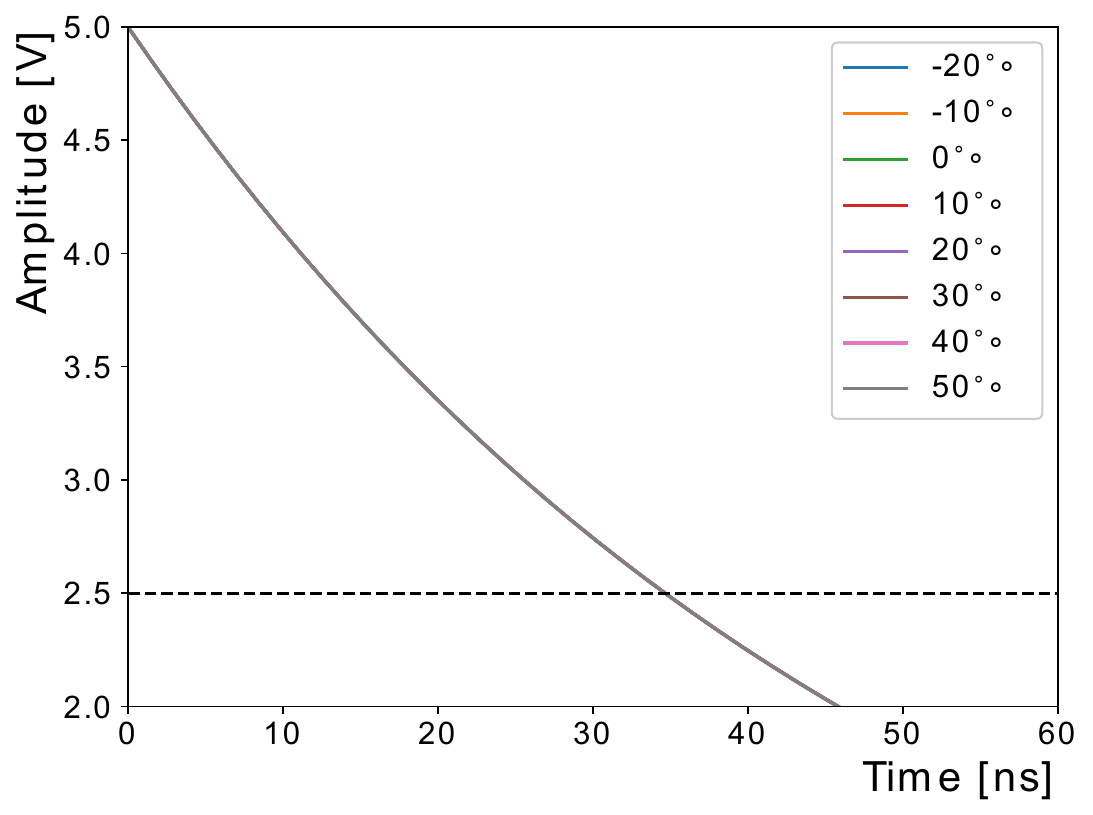}
    \caption{Temperature dependency of the shaping circuit when operating at temperatures ranging from -20$^{\circ}$C to +50$^{\circ}$C. The output voltage changes by less than 1$\%$/$^{\circ}$C when reaching the baseline voltage of 2.5\,V.}
    \label{fig:temp_dep}
\end{figure}

The saturation current is then defined as

\begin{equation}
    I_S = I_{diff} + I_{gen} = \alpha e^{-qV_J/kT} + \beta e^{-qV_J/2kT},
\end{equation}
where $\alpha$ and $\beta$ denote the strengths of the current contributions. The expression can be reduced to,

\begin{equation}
    I_S = \kappa e^{-qV_J/\eta kT},
\end{equation}
where the parameter $\eta$ varies between 1 and 2 depending on which current component dominates, the generation, or the diffusion current. 

Figure \ref{fig:temp_dep} shows that the output voltage of the shaper ($C=$1\,nF and $R=$50\,$\Omega$) varies by less than 1$\%$/$^{\circ}$C when operating at temperatures from -20$^{\circ}$C to 50$^{\circ}$C. Although the saturation current varies with temperature, the circuit performance is unaffected when a resistive load is connected and the diode reverse current increases by several orders of magnitude.

\section{Prototype circuit characterization}

\begin{figure}[h!]
    \centering
    \includegraphics[width=12cm]{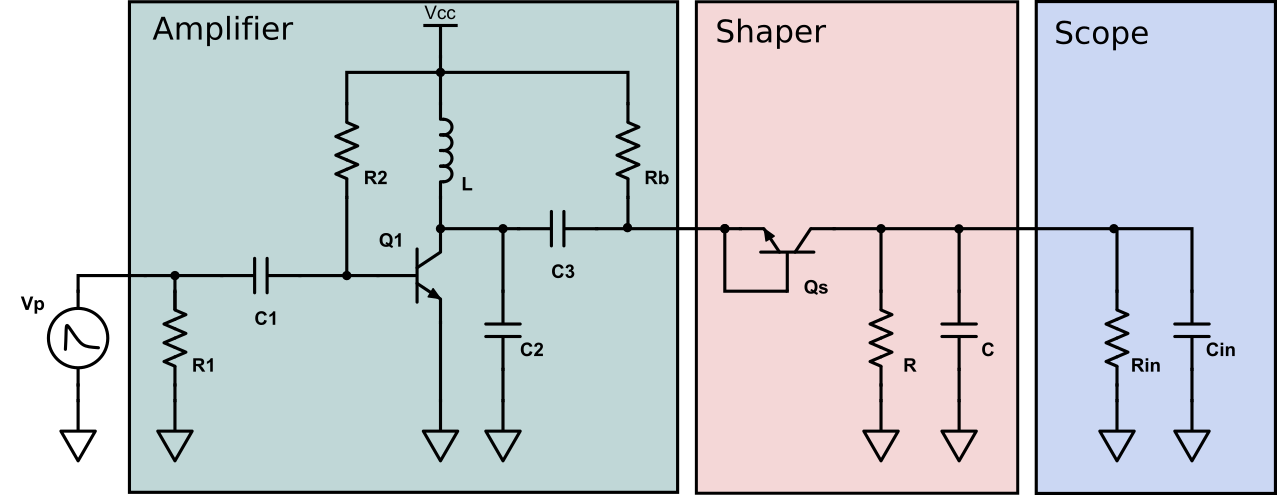}
    \caption{Composition of the characterization circuit: amplification stage (green), signal shaper (red), and oscilloscope (blue).}
    \label{fig:calib_circuit}
\end{figure}

As shown in Figure \ref{fig:calib_circuit}, we implemented a pulse shaper. The circuit consists of an amplification and shaping stage. The BFU760 bipolar transistor ($Q_s$) was set to diode mode by short-circuiting the base-emitter terminals. The capacitance of the transistor junctions meets the bandwidth requirements of the input signal.

The resistor $R_b$ sets a signal baseline of 1.1\,V and the shaper resistor $R$ was fixed at 1\,k$\Omega$. We kept the shaping capacitance $C$ as a free parameter to test different values. The amplification stage has a gain of $\sim$12. We fed the circuit with a photosensor-like signal generated using an Agilent 81106A signal generator to assess the dynamic range and linearity. The signal traces were recorded using a R$\&$S RTO-1044 sampling scope (2 GHz) with active, high-impedance probe head to minimize additional circuit load.

\begin{figure}[h!]
    \centering
    \includegraphics[width=8cm]{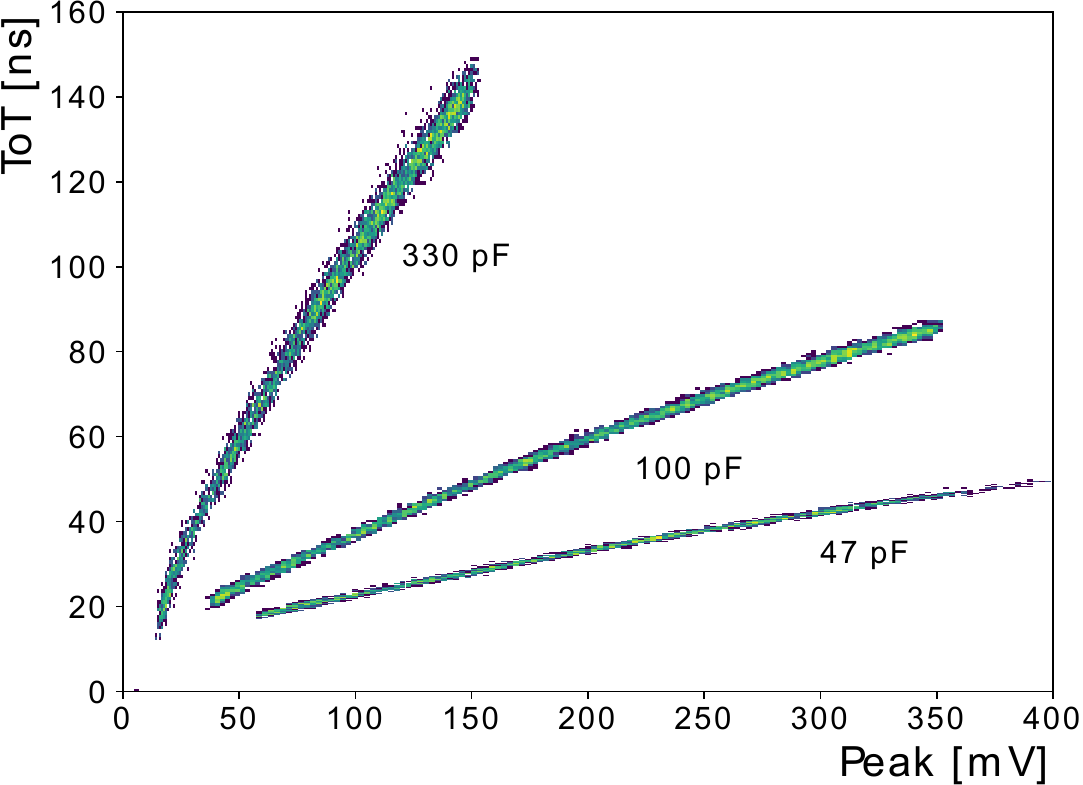}
    \caption{ToT dependency on the shaper capacitance. }
    \label{fig:cap_comparison}
\end{figure}

In the first circuit evaluation, the leading and trailing edge and the corresponding ToTs (assuming a constant threshold of 10\,mV) were evaluated offline from the stored scope traces. Their correlation with the input peak amplitude is shown in Fig.\,\ref{fig:cap_comparison}. Three capacitance values were evaluated: 330\,pF, 100\,pF, and 47\,pF. The input signal to the shaper ranges from 50\,mV to 500\,mV. The shaper capacitance acts as a low-pass filter, for $C =$330\,pF, the shaper output signal is attenuated to about a third of the input signal amplitude. The output ToT/amplitude ratio was 1.07 $\pm$ 0.108\,ns/mV. Decreasing $C$ to 100\,pF improved the ratio to 0.31 $\pm$ 0.074\,ns/mV, attenuating the input amplitude by $70\%$. The best performance was obtained with $C = 47$\,pF. With this capacitance, the ToT/amplitude ratio decreases to 0.18 $\pm$ 0.045\,ns/mV preserving $90\%$ of the signal amplitude.

\begin{figure}[h!]
    \centering
    \includegraphics[width=6.5cm]{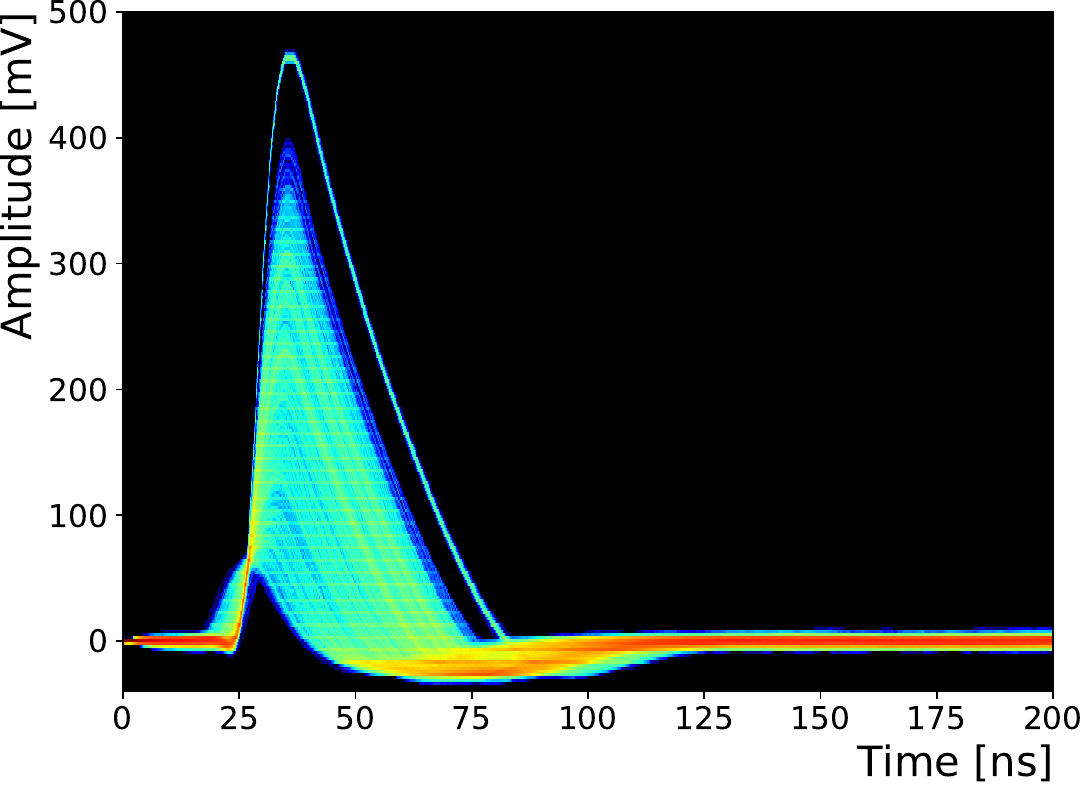}
    \includegraphics[width=6.5cm]{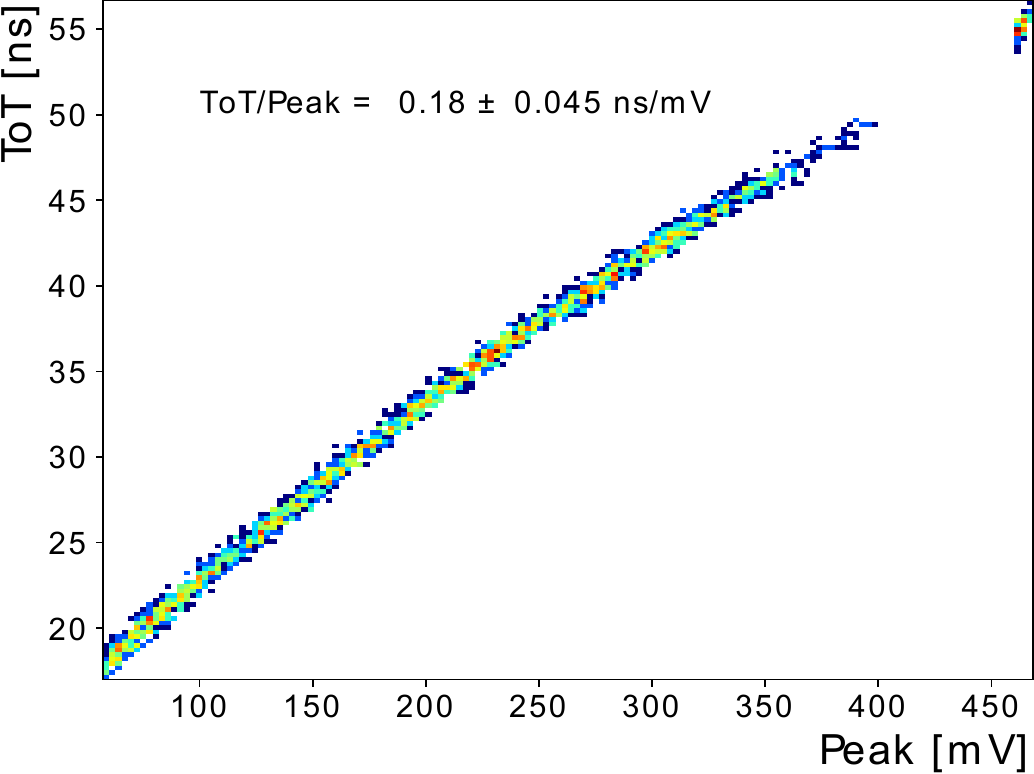}
    \caption{Output signal of the shaper circuit (left). Amplitude vs.\ ToT for $C = 47$\,pF and $R = 1$\,k$\Omega$. The characteristic slope is 0.18 $\pm$ 0.045\,ns/mV (right). The discrimination threshold for ToT evaluation was set to 10\,mV.}
    \label{fig:calibration}
\end{figure}

The subsequent analysis of the circuit was performed for $C = 47$\,pF. Figure \ref{fig:calibration} shows the shaped signal and the amplitude vs.\ ToT characteristic of the final design. We observed that the system maintains linearity throughout the amplitude range with an estimated resolution of $\sigma_{\frac{ToT}{Amp}}/\Delta_{\frac{ToT}{Amp}}= 4\%$.

\begin{figure}[h!]
    \centering
    \includegraphics[width=6.5cm]{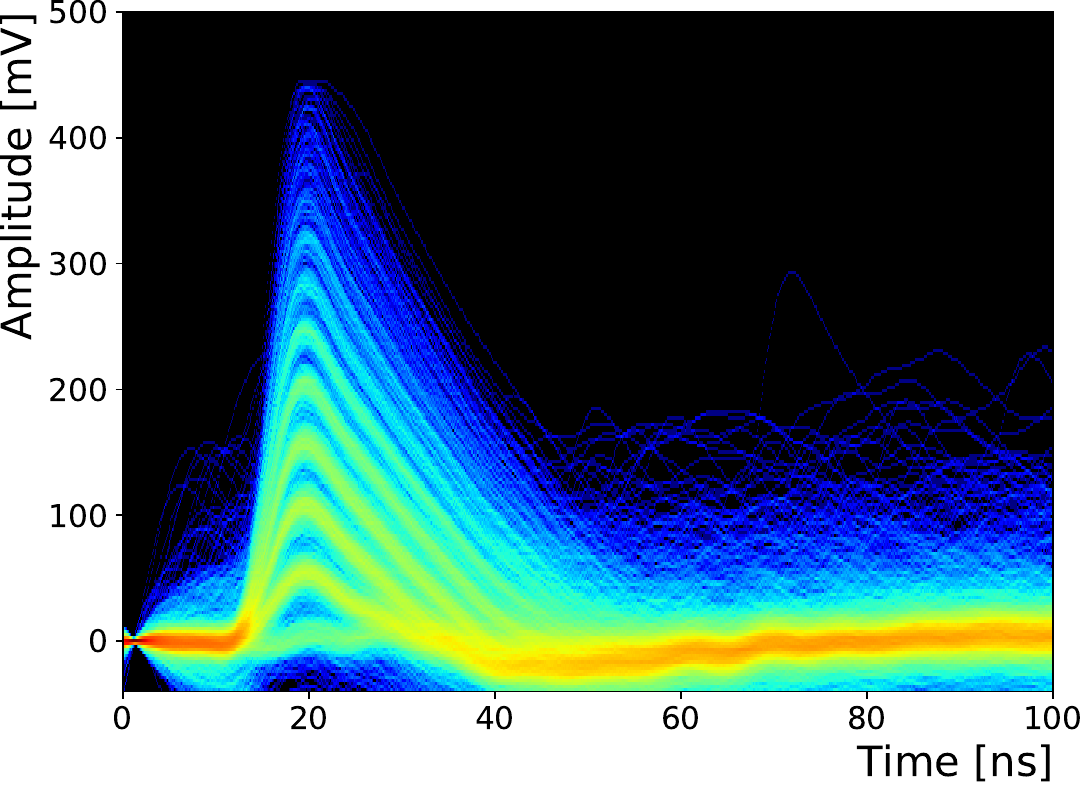}
    \includegraphics[width=6.5cm]{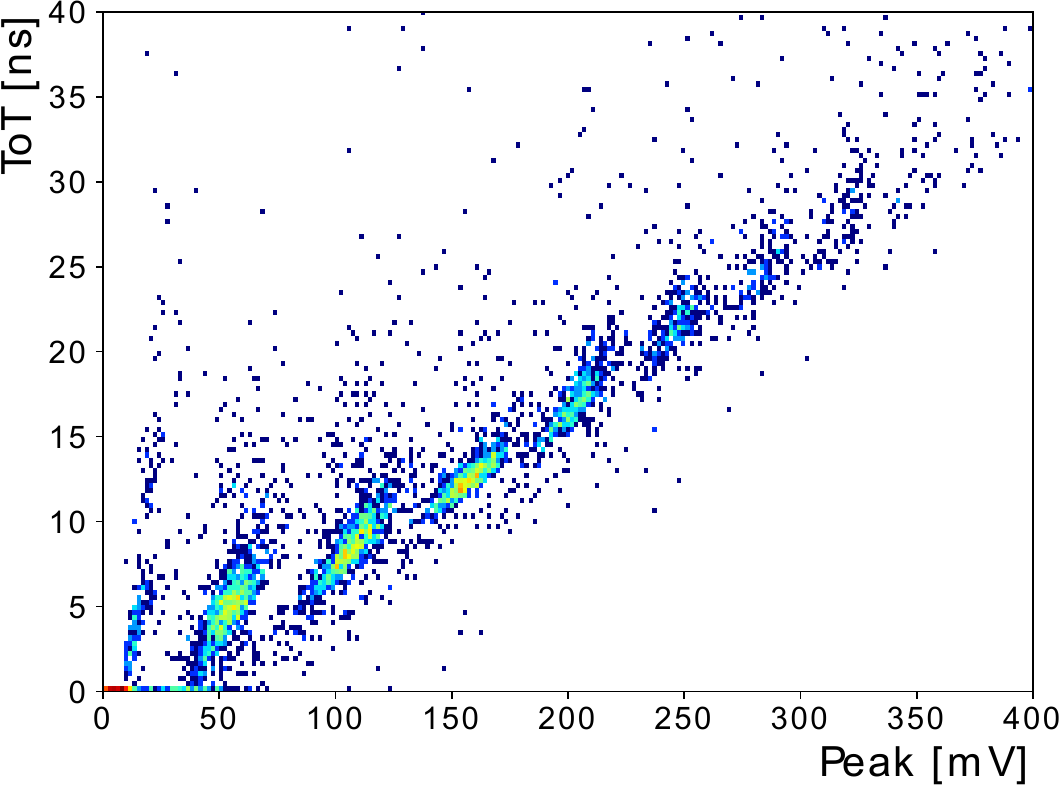}
    \caption{Signal shaping (left) and amplitude vs.\ ToT spectrum (right) are shown for the AFBR-S4N66P024M SiPM. The ToT threshold was set to 10\,mV.}
    \label{fig:sipm_test}
\end{figure}

We also tested the shaper circuit using real photon sensor signals from an AFBR-S4N66P024M SiPM. The SiPM was installed in a dark box where a picosecond laser pulser LD-405 ($405$\,nm/100\,kHz) illuminated it with light pulses of few photons per pulse, resulting in large statistical pulse-to-pulse fluctuations in photon count and corresponding signal amplitude. Figure \ref{fig:sipm_test} shows the SiPM signal after shaping and the amplitude vs.\ ToT spectrum. The amplitude of the SiPM signal ranges from $\sim 50$\,mV (1\,pe) to $\sim 450$\,mV. The measured ToT is linearly related to the signal amplitude, with the pe-ToT equivalent being $\sim$5\,ns.

\section{Implementation and testing in the CBM RICH DIRICH readout module}

The readout system of the Ring Imaging Cherenkov detector (RICH) of the Compressed Baryonic Matter (CBM) and the High Acceptance DiElectron Spectrometer (HADES) experiments is based on the DIRICH electronics \cite{Becker2023}.

\begin{figure}[h!]
    \centering
    \includegraphics[width=12cm]{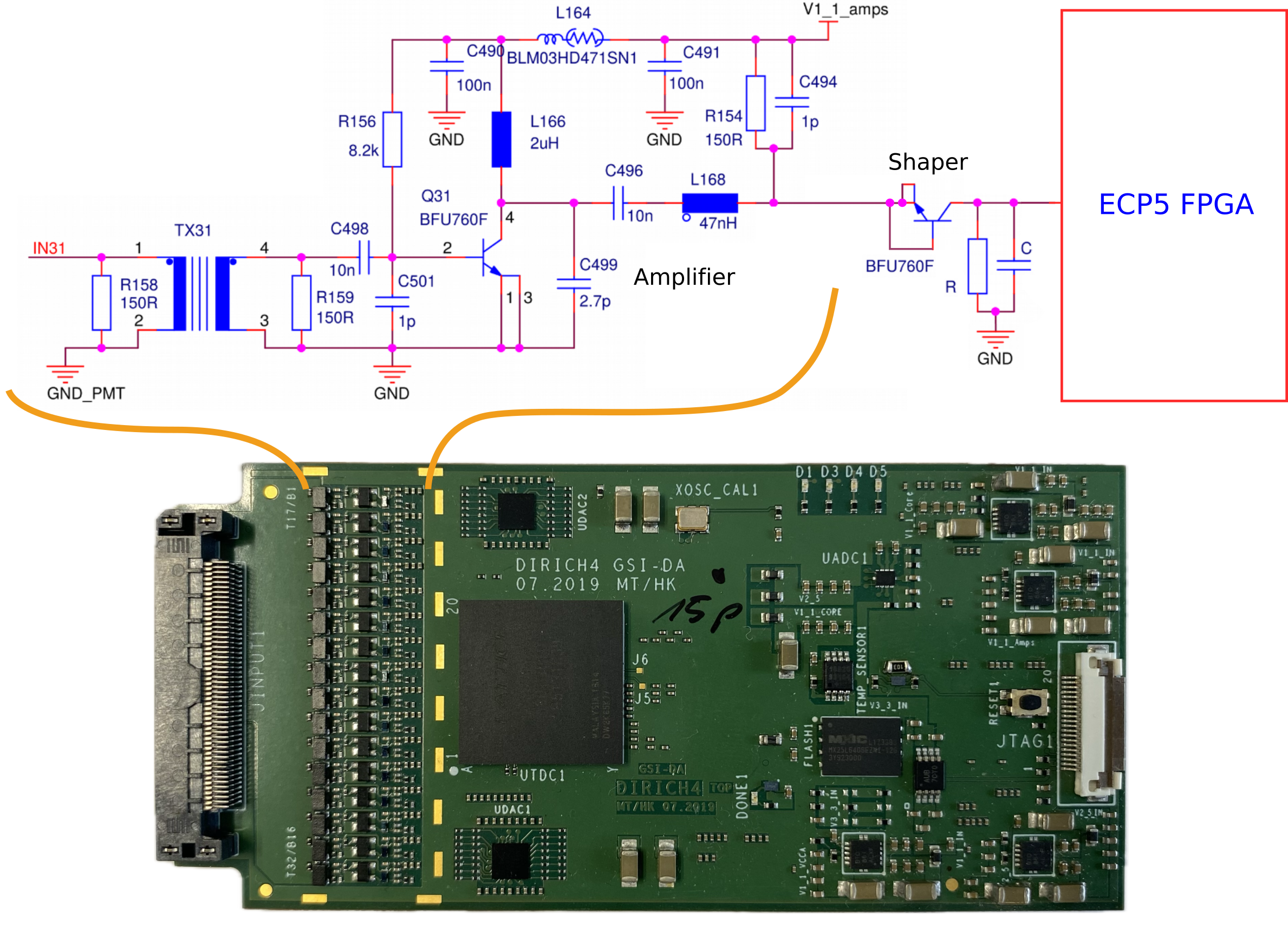}
    \caption{Pre-amplification stage of a DIRICH channel (top). DIRICH readout board (bottom).}
    \label{fig:dirich_daq}
\end{figure}

The DIRICH front-end module provides 32 individual channels of high-precision FPGA-TDC-based time measurement. Each channel includes a discrete transistor-based fast input amplifier with a gain of $\sim$ 30 and 4\,GHz bandwidth. Signal discrimination and the subsequent leading and trailing edge time measurements (as well as data readout) are implemented on the ECP5 FPGA, which provides timing precision better than 20\,ps (RMS). Figure\,\ref{fig:dirich_daq} shows the DIRICH electronics. The DIRICH TDC measures the leading and trailing edges of the signal above a constant threshold set by a 16-bits Delta-Sigma DAC (Digital-to-Analog Converter) with a resolution of $\sim 0.04$\,mV.

\begin{figure}[h!]
    \centering
    \includegraphics[width=10cm]{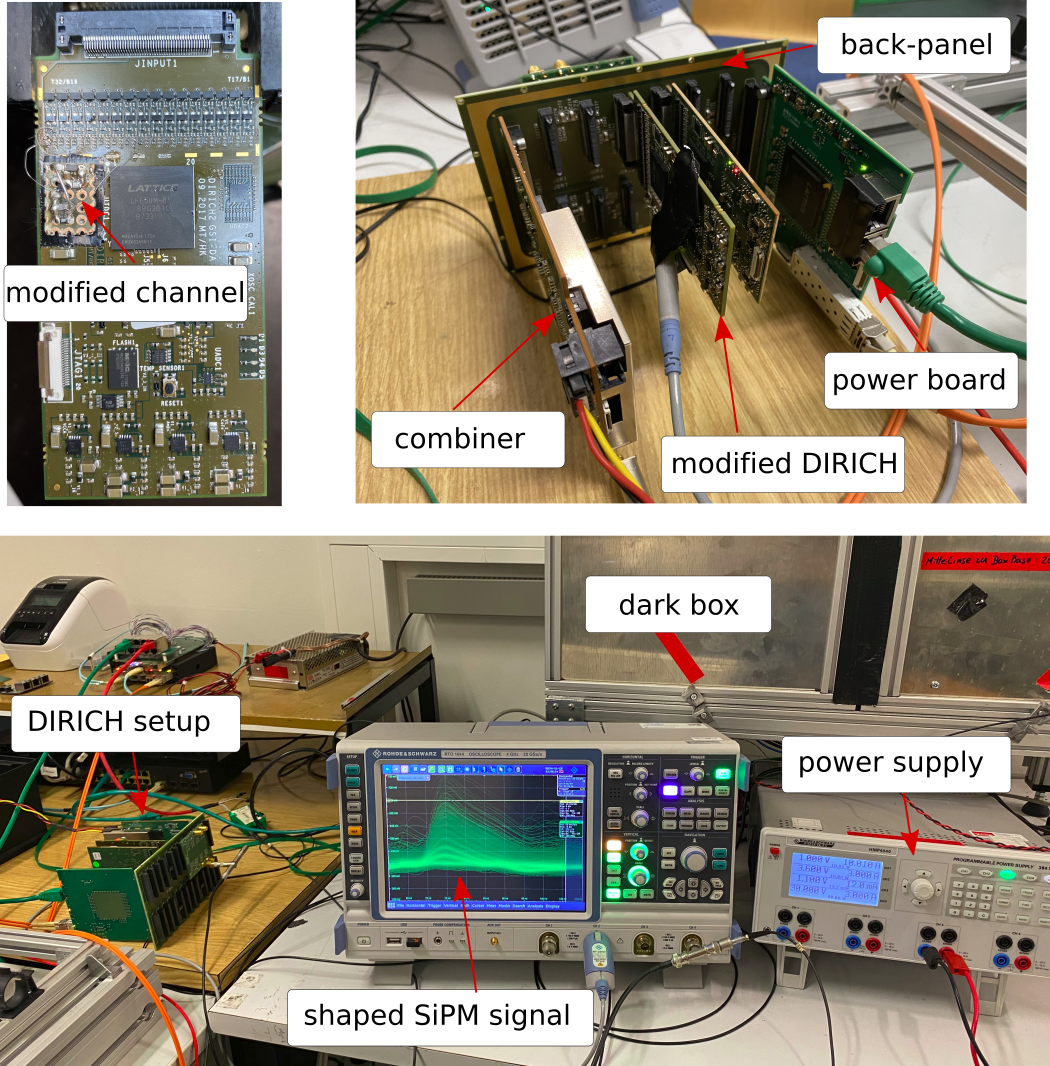}
    \caption{DiRICH setup for testing the amplitude-to-ToT system.}
    \label{fig:dirich_tot}
\end{figure}

We modified one of the DIRICH readout channels to evaluate the amplitude-to-ToT converter under real conditions. The test setup is shown in Fig.\,\ref{fig:dirich_tot}. The shaping circuit was mounted on the $31^{\rm st}$ DIRICH channel. The capacitor $C$ was set to 47\,pF and the resistor $R$ to 1\,k$\Omega$. The differential input impedance of the ECP5 FPGA is 100\,$\Omega$. 

The DIRICH board was installed in the back-panel module, where a power module powers the readout electronics, and a combiner module transmits the data to the central server.

\begin{figure}[h!]
    \centering
    \includegraphics[width=10cm]{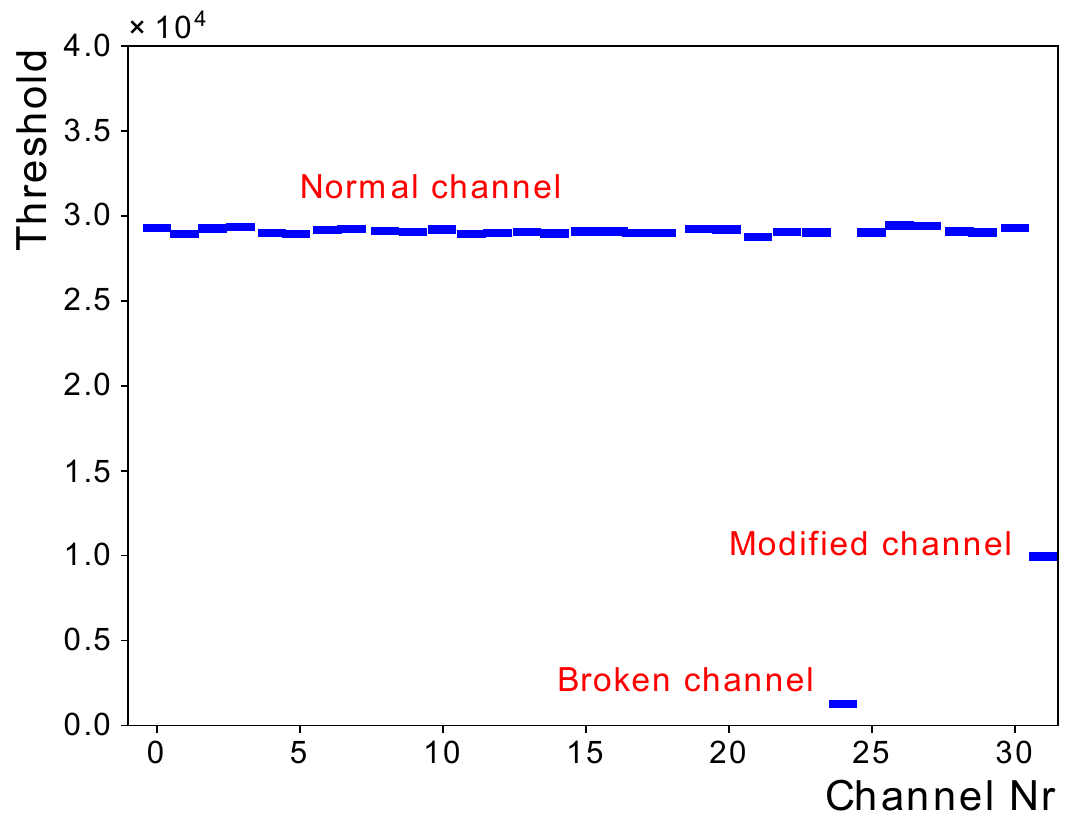}
    \caption{Baseline scanning of the DIRICH readout system. The baseline value for the standard ToT channels is around 1.1\,V (29\,250\,UDAC). The baseline for the modified $31^{\rm st}$ channel drops to 0.38\,V (9\,950\,UDAC). The $24^{\rm th}$ channel is a broken and does not participate in the test.}
    \label{fig:dirich_baseline}
\end{figure}

\begin{figure}[h!]
    \centering
    \includegraphics[width=10cm]{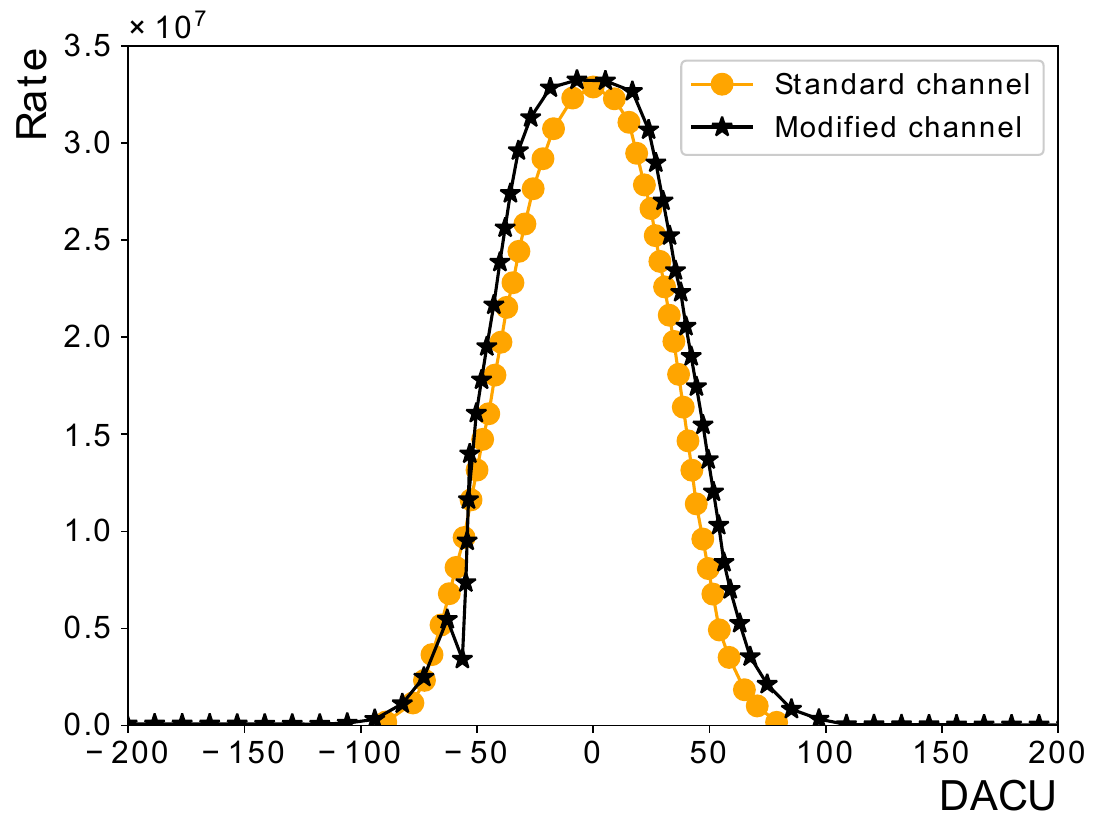}
    \caption{Baseline distribution of a normal (circles) and modified (stars) channel. In both cases, the baseline standard deviation is around 40\,UDAC.}
    \label{fig:dirich_baseline_var}
\end{figure}

We performed a channel baseline scan to compare the modified channel to a standard DIRICH channel. Figures\,\ref{fig:dirich_baseline} and \ref{fig:dirich_baseline_var} illustrate the results of the scan. The DIRICH establishes a  baseline around 29\,250\,UDAC (1.1\,V). for the modified channel, the baseline drops to 9\,950\,UDAC (0.38\,V) taking into account that the voltage drop between the base-collector diode of the BFU760F is $\sim$ 0.72\,V. We also compared the baseline variation. The modified ToT channel slightly increases the typical baseline variation, which follows a Gaussian distribution with a standard deviation of $\sim$40\,UDAC.

\begin{figure}[h!]
    \centering
    \includegraphics[width=10cm]{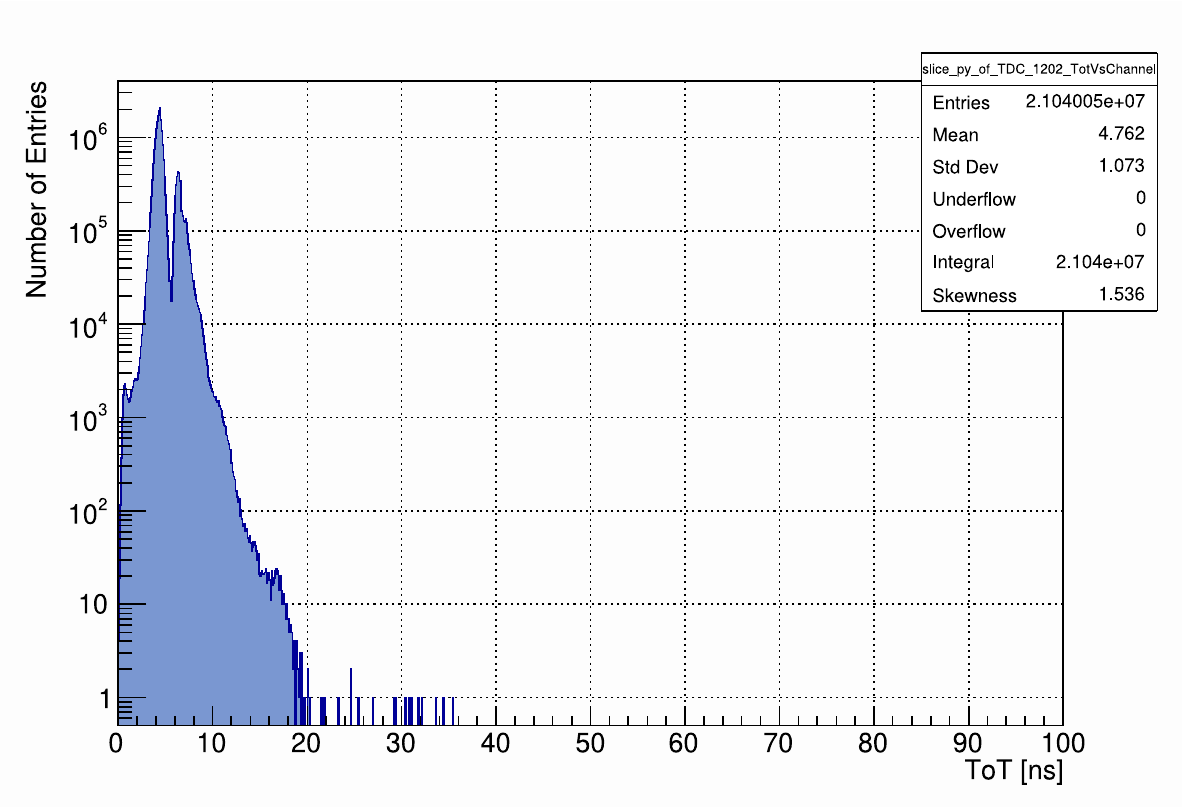}
    \includegraphics[width=10cm]{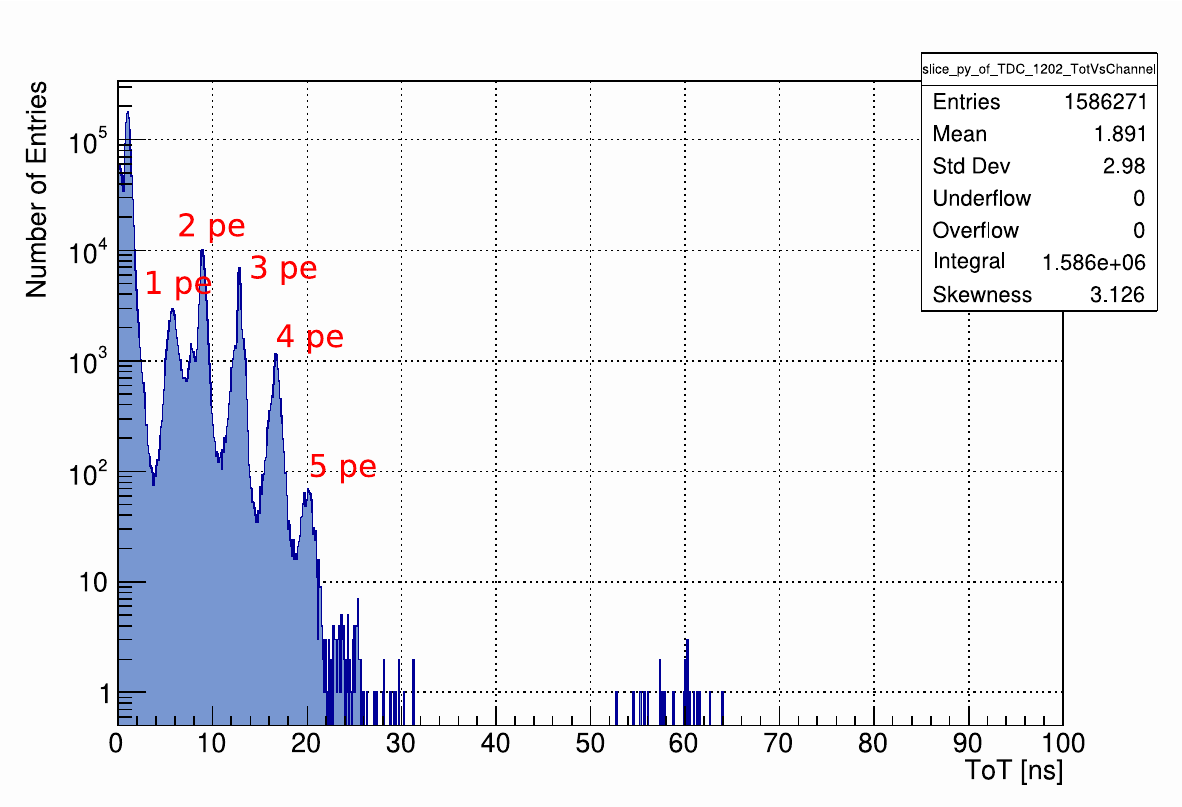}
    \caption{ToT spectrum of a standard (top) and a modified (bottom) DIRICH channel.}
    \label{fig:dirich_channel}
\end{figure}

We fed signals from the AFBR-S4N66P024M SiPM operating in dark conditions into the modified $31^{\rm st}$ ToT channel. The SiPM bias was set to 40\,V, and an amplifier increases the signals by a factor of $\sim$12. Figure\,\ref{fig:dirich_channel} shows a comparison of the ToT spectrum of a common DIRICH channel and the modified one.

The standard DIRICH channel cannot distinguish between the different levels of photoelectrons in the SiPM signal due to the exponential trailing edge. However, the modified channel can distinguish among the levels. The ToT equivalent to one photoelectron is $\sim$4\,ns.

\section*{Conclusions}

We present a new shaping circuit that improves the linearity, bandwidth, and dynamic range of single-threshold ToT systems. Composed of three passive elements -- a fast diode, a capacitor, and a resistor -- the circuit provides low power consumption and high integrability, favoring its implementation in ASIC technology. 

We performed Spice simulations of the shaping circuit using two diodes (1N5817 and SMS7621) and a BFU760 transistor connected in diode mode. We observed that junction capacitance ($C_{JO}$) and parasitic resistance ($R_S$) play a crucial role in shaping performance. The characteristic amplitude vs.\ ToT of the 1N5817 ($C_{JO} = 190$\,pF and $R=0.05$\,$\Omega$) showed a linear trend. In contrast, the characteristics of the SMS7621 ($C_{JO} = 0.1$\,pF and $R=12$\,$\Omega$) was nonlinear when using a shaping capacitance of 1\,nF and a resistive load of 50\,$\Omega$. The shaper linearity of the SMS7621 and BFU760 improved after bandwidth broadening by decreasing the shaping capacitance and increasing the resistive load to 47\,pF and 1\,k$\Omega$.

Implementing the circuit revealed that increasing the shaping capacitance $C$ worsened the ToT resolution. We evaluated three capacitance values: 330\,pF, 100\,pF, and 47\,pF, achieving ToT resolutions of 10$\%$, 4.1$\%$, and 4$\%$ respectively. The circuit was tested under real conditions by feeding it with signals from an AFBR-S4N66P024M SiPM. This showed a linear relationship between the photoelectron level and the ToT of 5\,ns/pe.

The shaping circuit was also tested in a ToT channel of the DIRICH electronics readout of the CBM RICH detector. We compared the baseline and ToT spectra of standard and modified DIRICH channels and found that the baseline of the modified channel decreases from 1.1\,V to 0.38\,V due to the diode voltage ($\sim$ 0.7\,V). Despite this change, the baseline standard deviation remained around 40\,UDAC. However, unlike the standard DIRICH channels, the modified one differentiates between the different photoelectron levels of the SiPM signal, providing a ToT/pe equivalent of $\sim$4\,ns.\\

\subsection*{Acknowledgments} \noindent This work has been supported by ``Netzwerke 2021'', an initiative of the Ministry of Culture and Science of the State of North Rhine-Westphalia.

\bibliographystyle{elsarticle-num} 
\bibliography{SiPM_fram}






\end{document}